\documentclass[twocolumn, aps, prb]{revtex4}

\usepackage{graphicx}
\usepackage{dcolumn}
\usepackage{bm}

\begin{document}

\title{Molecular dynamics simulations of thermal conductivity and spectral phonon relaxation time in suspended and supported graphene}

\author{Bo Qiu}
\author{Xiulin Ruan}%
 \email{ruan@purdue.edu}
\affiliation{%
School of Mechanical Engineering and the Birck Nanotechnology Center\\
    Purdue University\\
    West Lafayette, Indiana 47907-2088\\
}%

\date{\today}

\begin{abstract}

We perform molecular dynamics (MD) simulations with phonon spectral analysis aiming at understanding the two dimensional (2D) thermal transport in suspended and supported graphene. Within the framework of equilibrium MD simulations, we perform spectral energy density (SED) analysis to obtain the lifetime of individual phonon modes. The per-mode contribution to thermal conductivity is then calculated to obtain the lattice thermal conductivity in the temperature range 300-650 K. In contrast to prior studies, our results suggest that the contribution from out-of-plane acoustic (or ZA) branch to thermal conductivity is around 25-30$\%$ in suspended single-layer graphene (SLG) at room temperature. The thermal conductivity is found to reduce when SLG is put on amorphous SiO$_2$ substrate. Such reduction is attributed to the strengthened scattering in all phonon modes in the presence of the substrate. Among them, ZA modes are mostly affected with their contribution to thermal conductivity reduced to around 15$\%$. As a result, thermal transport is dominated by in-plane acoustic phonon modes in supported SLG.

\end{abstract}


\maketitle

\section{Introduction}
Ever since its experimental isolation in 2004 \cite{Novoselov_2004}, graphene has been the focus of extensive studies because of a number of fascinating properties it exhibits \cite{Bolotin_2008,Lee_2008,Balandin_2008}. Aside from high electronic mobility \cite{Geim_2007}, 2D thermal transport and superior thermal conductivity have also been demonstrated in graphene experimentally \cite{Balandin_2008, Seol_2010}. It was suggested that such 2D transport can lead to infinitely large intrinsic thermal conductivity \cite{Balandin_2011,Basile_2006, Cai_2010}. Also, it leads to novel three dimensional (3D) to 2D crossover phenomena in thermal conductivity when the material dimensions shrink from corresponding bulk \cite{Ghosh_2010,Qiu_2010}. In addition, when extrinsic perturbations, for instance, corrugation, grain boundaries, substrate, etc., are present, additional phonon scattering channels can be opened and contribute to the thermal resistance in a manner distinguished from those typically found in other material systems.

Despite extensive investigations on the electronic transport in graphene, many issues of thermal transport still remain unclear. For instance, reported thermal conductivity values of suspended SLG spread widely between 1,800 and 5,300 W/m-K at room temperature \cite{Balandin_2008, Cai_2010, Lee_2011} possibly due to the varying degrees of sample quality, rendering it controversial in identifying 2D thermal transport features in SLG. In order to address these issues, many theoretical approaches have been proposed \cite{Hu_2009, Guo_2009, Lindsay_2010,Tan_2011,Nika_2009}. Among them, Nika $\emph{et al}$ \cite{Nika_2009} suggested negligible contribution from ZA phonons due to their low group velocity. Lindsay $\emph{et al}$ \cite{Lindsay_2010} suggested, on the contrary, that the majority heat is carried by ZA phonons. It is evident that agreement on the relative importance of individual phonon modes to the thermal conductivity in suspended graphene has not been reached, largely because these models used different assumptions that cannot be verified directly with experiments. This issue extends to the understanding of the reduced thermal conductivity observed in supported graphene \cite{Seol_2010} while few quantitative theoretical models \cite{Ong_2011} without adjustable parameters have been developed to predict its thermal conductivity so far.

In this work, we employ MD simulations in combination with SED analysis to extract spectral phonon lifetime, mean free path (MFP) and thermal conductivity for both suspended and supported SLG in the temperature range 300-650 K. Contrary to previous suggestions, we find that the out-of-plane phonons (ZA and ZO) couple to the in-plane phonons in suspended SLG with medium strength, and the contribution of ZA phonon to the total thermal conductivity is neither negligible nor dominant. We also find that the contributions from all phonon modes to thermal conductivity are reduced in the presence of substrate while ZA modes are mostly affected.

\section{Methodology}

\subsection{Molecular dynamics simulations setup}

\begin{figure}
  \begin{center}
    \includegraphics{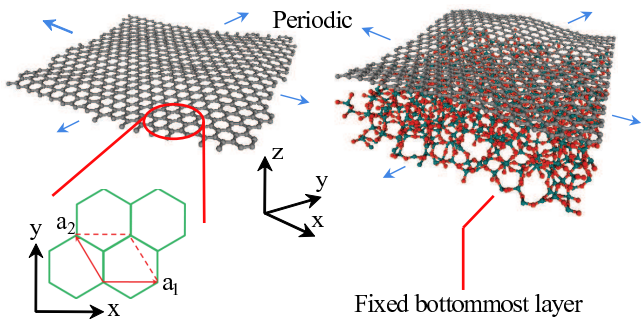}
  \end{center}
  \caption{(Color online) Instantaneous geometry of both suspended and supported SLG. The SLG plane is parallel to the x-y plane while the SLG out-of-plane direction aligns with z.}
 \label{fig:structure}
\end{figure}

In Fig. \ref{fig:structure} we present the MD simulation domain setup for both suspended and supported SLG. The dimensions of the shown domains are $4.4\times 4.3 \times 1.6$ nm$^3$ for length, width, and height, respectively. Periodic boundary condition (PBC) is applied in the in-plane directions (parallel to graphene plane) while no restriction is applied to SLG vibration in the out-of-plane direction. We choose silicon dioxide (SiO$_2$) as the substrate, which is usually used in experiments. A 2 nm thick block of amorphous SiO$_2$ which comply with the PBC of the graphene domain is prepared by following the heating-quenching recipe used by Ong \emph{et al} \cite{Ong_2010}. Similar thickness of substrate has been successfully used previously for both electronic \cite{Miwa_2005, Vallabhaneni_2011, Orellana_2008} and thermal simulations \cite{Ong_2010, Ong_2011}. Thicker SiO$_2$ block is not found to induce obvious difference in the in-plane thermal transport of supported graphene. To ensure stability and to approximate semi-infinite substrate, the bottom layer of amorphous SiO$_2$ is fixed. The graphene sheet is released from 2 $\AA$ above the substrate and allowed to conform to the SiO$_2$ surface freely.  The geometries of both suspended and supported graphene are pre-optimized under constant pressure and temperature (NPT) for 6 ns to ensure zeroed pressure, equilibrated temperature and stablized interfacial structure before being used for individual simulation runs.

The LAMMPS package is used to perform all MD simulations \cite{Plimpton_1995}. Optimized Tersoff (OPT) potentials \cite{Lindsay_2010_1} are adopted to model C-C interactions in SLG. Interactions within SiO$_2$ are modeled using Tersoff potentials parameterized by Munetoh \emph{et al} \cite{Munetoh_2007}. C-Si and C-O interactions are assumed to be of van der Waals (vdW) type \cite{Miwa_2005} and modeled using Lennard-Jones (LJ) potentials
\begin{equation}
V(r_{ij})=
4\epsilon[(\frac{\sigma}{r_{ij}})^{12}-(\frac{\sigma}{r_{ij}})^{6}]
\label{eqn:lj}
\end{equation}
with parameters $\epsilon_{C-Si}=8.909$ meV, $\sigma_{C-Si}=3.326$ $\AA$, $\epsilon_{C-O}=3.442$ meV and $\sigma_{C-O}=3.001$ $\AA$ \cite{Rappe_1992}. The cutoff is chosen to be $2.7\sigma$. Here $r_{ij}$ is the separation between atom $i$ and $j$. The timesteps for suspended and supported SLG simulations are chosen to be 0.8 and 0.2 fs, respectively, to ensure stability and resolution of all possible vibrational frequencies. Starting from pre-optimized geometries, after initial equilibration in constant volume and temperature ensemble (NVT), the systems are run in the constant volume and energy ensemble (NVE), from which the atomic velocities are extracted and post-processed. The phonon properties are found to converge for simulation lengths of 3.2 and 1.0 ns in NVE ensemble for suspended and supported SLG, respectively. In this work, only the thermal transport in x direction (zig-zag) is considered. Five independent simulations runs are performed and averaged for each case study to minimize effects from statistical fluctuations. Since the temperatures at which we perform the MD simulations are below the Debye temperature of SLG ($\sim$ 1000-2300 K \cite{Tewary_2009}), quantum corrections (QC) outlined in the Appendix have been performed, which were successfully practiced in similar ways in previous works \cite{Che_2000, Hu_2009}.

\subsection{Spectral energy density approach}

In most cases, the intrinsic phonon transport in crystals is dominated by the three-phonon Normal or Umklapp process. Therefore, considering only up to third-order anharmonic force constants is usually a good approximation. However, in CNT's, neglecting higher-order anharmonic interactions is found to significantly overestimate phonon lifetime \cite{Thomas_2010}. Due to the close connection between CNT and SLG, we expect similar overestimation in SLG if a fully anharmonic simulation is not performed. To predict phonon dispersion relations and lifetime of SLG, we employ the phonon SED analysis which, in combination with MD simulations, naturally includes the fully temperature-dependent anharmonicity of the atomic interactions. Such analysis and similar formulations have been successful in predicting individual phonon properties for various material systems \cite{Ladd_1986, McGaughey_2004, Turney_2009, Henry_2008, Koker_2009, Shiomi_2006}, and hence will be used here.

Although phonon lifetime prediction using only atomic velocity data was reported in recent SED works on CNT \cite{Thomas_2010}, PbTe \cite{Qiu_2011_pbte}, suspended graphene \cite{Qiu_2011_mrs, Qiu_2011_imece}, and supported CNT \cite{Ong_2011_cnt}, it was later pointed out by Larkin \emph{et al} that eigen-displacements should also be included in the formulation of SED functions \cite{Larkin_2012}. A detailed derivation of the connection between SED function and phonon spectral properties was given in Ref. \cite{Larkin_2012}. Here we present a simplified version to illustrate the methodology.

In a harmonic system, phonons do not scatter. Therefore they have infinite lifetimes and exhibit delta-function type peaks in vibrational spectrum. In real material systems, phonons scatter and thus have finite lifetimes due to the anharmonic interactions, leading to shift in the phonon spectrum and broadening of the phonon peaks. According to Ladd \emph{et al} \cite{Ladd_1986}, the normal mode amplitude can be written under single mode relaxation time approximation as
\begin{eqnarray}
S_{\mathbf{k},\nu}(t)=S_{\mathbf{k},\nu,0} e^{-i(\omega_{\mathbf{k},\nu}^A-i\Gamma_{\mathbf{k},\nu})t}.
\label{eqn:ladd}
\end{eqnarray}
Here $S_0$ is the magnitude of the vibration, $\omega^A$ is the anharmonic angular phonon frequency, $\Gamma$ is the phonon linewidth, $\mathbf{k}$ is the wavevector, $\nu$ is the index of phonon branches. Then Fourier transform of the time derivative of normal mode amplitude is
\begin{eqnarray}
F[\dot{S}_{\mathbf{k},\nu}(t)]=\frac{1}{\sqrt{2\pi}}\int_{-\infty}^{\infty}S_{\mathbf{k},\nu,0} (-i\omega_{\mathbf{k},\nu}^A-\Gamma_{\mathbf{k},\nu}) \nonumber \\ \times e^{-i(\omega_{\mathbf{k},\nu}^A-\omega)t}e^{-\Gamma_{\mathbf{k},\nu}t}dt.
\label{eqn:ftqdot}
\end{eqnarray}
Its norm square is:
\begin{eqnarray}
|F[\dot{S}_{\mathbf{k},\nu}(t)]|^2=\frac{S_{\mathbf{k},\nu,0}^2((\omega_{\mathbf{k},\nu}^A)^2+\Gamma_{\mathbf{k},\nu}^2)/2\pi}{(\omega-\omega_{\mathbf{k},\nu}^A)^2+\Gamma_{\mathbf{k},\nu}^2}.
\label{eqn:modftqdot}
\end{eqnarray}
This function is in the Lorentzian form with its full width at half maximum (FWHM) equals $2\Gamma$. In addition, the phonon lifetime $\tau$ is directly related to the linewidth as \cite{Ladd_1986}:
\begin{eqnarray}
\tau=1/2\Gamma.
\label{eqn:tau}
\end{eqnarray}
Therefore, when SED function is defined as
\begin{eqnarray}
\Psi(\mathbf{k},\nu,f)&\equiv&|F[\dot{S}_{\mathbf{k},\nu}(t)]|^2 \nonumber \\
&=&\frac{S_{\mathbf{k},\nu,0}^2((\omega_{\mathbf{k},\nu}^A)^2+\Gamma_{\mathbf{k},\nu}^2)/2\pi\Gamma_{\mathbf{k},\nu}^2}{[4\pi\tau_{\mathbf{k},\nu}(f-f_{\mathbf{k},\nu}^A)]^2+1} \nonumber \\
&\equiv&\frac{C_{\mathbf{k},\nu}}{[4\pi\tau_{\mathbf{k},\nu}(f-f_{\mathbf{k},\nu}^A)]^2+1},
\label{eqn:sed}
\end{eqnarray}
the fully-anharmonic phonon frequency $f_{\mathbf{k},\nu}^A$ and lifetime $\tau_{\mathbf{k},\nu}$ associated with a particular phonon mode ($\mathbf{k},\nu$) can be extracted by fitting to the Lorentzian peak shape of SED function $\Psi(\mathbf{k},\nu,f)$. Here $f=\omega/2\pi$ is the phonon frequency.

\begin{figure}
  \centerline{\includegraphics{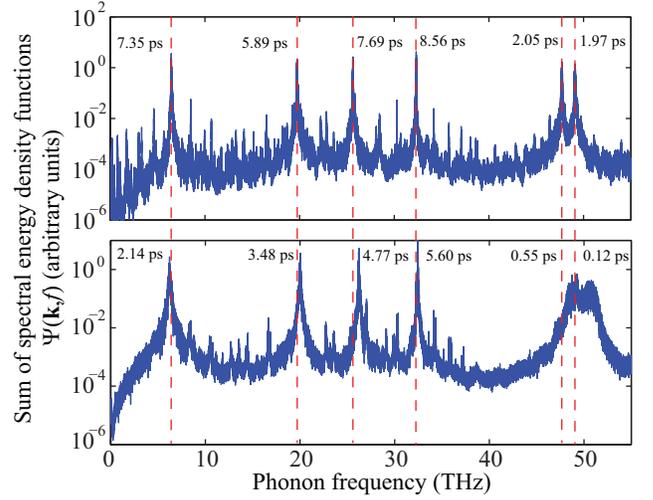}} \vspace{2mm}
  \caption{(Color online) Semilogarithmic plots of the sum of SED functions $\Psi(\mathbf{k},f)=\sum_{\nu}\Psi(\mathbf{k},\nu,f)$ along $\Gamma$-K direction with $\mathbf{k}=\mathbf{k}_{max}/2$. a) suspended SLG. b) supported SLG. The red dashed lines indicate phonon peak positions of suspended SLG. The numbers adjacent to each peak are lifetime values obtained through fitting to Eq. (\ref{eqn:sed}).}
  \label{fig:sed}
\end{figure}

On the other hand, the time derivative of the normal mode amplitude of mode ($\mathbf{k},\nu$) is given as \cite{Dove_1993}:
\begin{eqnarray}
\dot{S}_{\mathbf{k},\nu}(t)=\sum_{\alpha,b,l}^{3,n_b,N_c}\sqrt{\frac{m_b}{N_c}}\dot{u}_{\alpha}^{b,l}(t)e^{b*}_{\alpha}(\mathbf{k},\nu)\exp(-i\mathbf{k}\cdot \mathbf{r}_0^l)
\label{eqn:nm}
\end{eqnarray}
where $\alpha$ represents $x,y,z$ directions, $b$ is the index of basis atoms, $n_b$ is the number of basis atoms in the chosen cell, $l$ is the index of cells, $N_c$ is the total number of cells in the MD domain, and $m_b$ is the atomic mass of basis atom $b$.
$\dot{u}_{\alpha}^{b,l}$ is the $\alpha$ component of the velocity
of basis atom $b$ in cell $l$ and $\mathbf{r}_0^{l}$ is the
equilibrium position of cell $l$. $e^{b*}_{\alpha}(\mathbf{k},\nu)$ is the complex conjugate of the $\alpha$ component of the eigen-displacement of basis atom $b$ associated with the $\nu$th branch at a chosen wavevector $\mathbf{k}$. Based on the time history of atomic velocities generated by MD simulations and eigen-displacements from lattice dynamics (LD) calculations, the normal mode amplitude can be obtained according to Eq. (\ref{eqn:nm}). Then SED function can be constructed and fitted with Eq. (\ref{eqn:sed}) to extract $f_{\mathbf{k},\nu}^A$ and $\tau_{\mathbf{k},\nu}$, as shown in Fig. \ref{fig:sed}.

As mentioned in previous section, to minimize statistical fluctuations due to MD simulations, five independent simulations are run for each case study. The arithmetic average of the corresponding SED functions are found to give better Lorentzian peak shapes. Therefore, all data in this work are extracted from the averaged SED functions. However, it should be pointed out that artificial broadening of the peaks after averaging can be non-negligible if the peak positions from individual runs differ much. Such broadening will lead to underestimated phonon lifetime. To examine such effect, we assume two identical peaks are separated by a small amount $2\delta$. When they are super-positioned, the new peak centers at $\bar{f}$. Therefore, the positions where half maximum of the new peak occurs are solutions to
\begin{eqnarray}
\frac{1}{(f-\bar{f}-\delta)^2+\Gamma^2}+\frac{1}{(f-\bar{f}+\delta)^2+\Gamma^2}=\frac{1}{\delta^2+\Gamma^2}.
\label{eqn:stdv}
\end{eqnarray}
The new FWHM is then found as:
\begin{eqnarray}
2\Gamma'=2\Gamma\sqrt{2(\frac{\delta}{\Gamma})^2+\sqrt{5(\frac{\delta}{\Gamma})^4+2(\frac{\delta}{\Gamma})^2+1}}.
\label{eqn:twog}
\end{eqnarray}
In present study, it is usually observed that $\delta/\Gamma<0.4$ even for low frequency acoustic phonons. Therefore, the underestimation of phonon lifetime is expected to be well below 20$\%$. This is confirmed by the agreement between the average of phonon lifetimes extracted from individual runs and that from averaged SED function.

The thermal conductivity $\kappa_l$ can be obtained as a summation of contributions from all phonon modes in the full Brillouin zone (BZ):
\begin{equation}
\kappa_{l,x}=\sum_{\mathbf{k}}\sum_{\nu}c_v(\mathbf{k},\nu) v_{g,x}^2(\mathbf{k},\nu) \tau(\mathbf{k},\nu),
\label{eqn:kl}
\end{equation}
where $c_v$ is the quantum per-mode volumetric specific heat of a harmonic oscillator, which decreases rapidly as frequency increases:
\begin{equation}
c_v=\frac{k_B x^2 \exp(x)}{V[\exp(x)-1]^2}.
\label{eqn:cv}
\end{equation}
Here $V$ is the effective volume of the simulation domain, and $x=\hbar\omega/k_B T$. $v_g=d\omega/dk$ is the phonon group velocity.  The summation goes over all resolvable wavevector $\mathbf{k}$'s and all phonon branches $\nu$.

One limitation of SED approach is that only a number of $\mathbf{k}$ points which comply with the periodicity of the MD domain can be resolved, leading to the finite resolution of the $\mathbf{k}$-grid and cutoff of contribution from very long wavelength phonon modes \cite{Henry_2009, Turney_2009, Ong_2011_cnt}. The exclusion of these modes may lead to domain size effect of the thermal conductivity. Such effect generally presents in MD simulations with PBC specified rather than only exists in SED approach. However, due to the large amount of resolvable phonon modes ($\mathbf{k},\nu$) in BZ associated with larger domains, study of domain size effect is not as feasible for SED approach. Nonetheless, a reasonably sized simulation domain should preserve the validness of at least the qualitative description of physical processes.

One source of inaccuracy in our calculation may come from the fact that normal and Umklapp processes can both contribute to the broadened phonon peak in a non-distinguishable manner in MD simulations. While Umklapp process is directly resistive for thermal transport, normal process is only indirect - it increases the population of high wavevector phonons which can later encounter Umklapp process and lead to thermal resistance. Compared to the simple addition of these two processes in our SED method, a more accurate treatment would be a consideration of these two processes separately and a solution of the full Boltzmann transport equation \cite{Lindsay_2010, Singh_2011}. However, that approach cannot take into account higher order phonon scattering processes or extrinsic scattering mechanisms such as substrate effects which are important for our problem. Considering the fact that a high wavevector phonon generated from a normal process will likely participate in the Umklapp process at a later time and contribute to thermal resistance, the direct inclusion of normal process in lifetime calculation should be approximately valid. Also, the lifetime calculated from MD has already included the effects of non-equilbrium fluctuation of phonon distribution function, similar in spirit to that in the GK method \cite{McQuarrie_2000}.

\section{Results}

\subsection{Phonon lifetime}

\begin{figure}
\centerline{\includegraphics{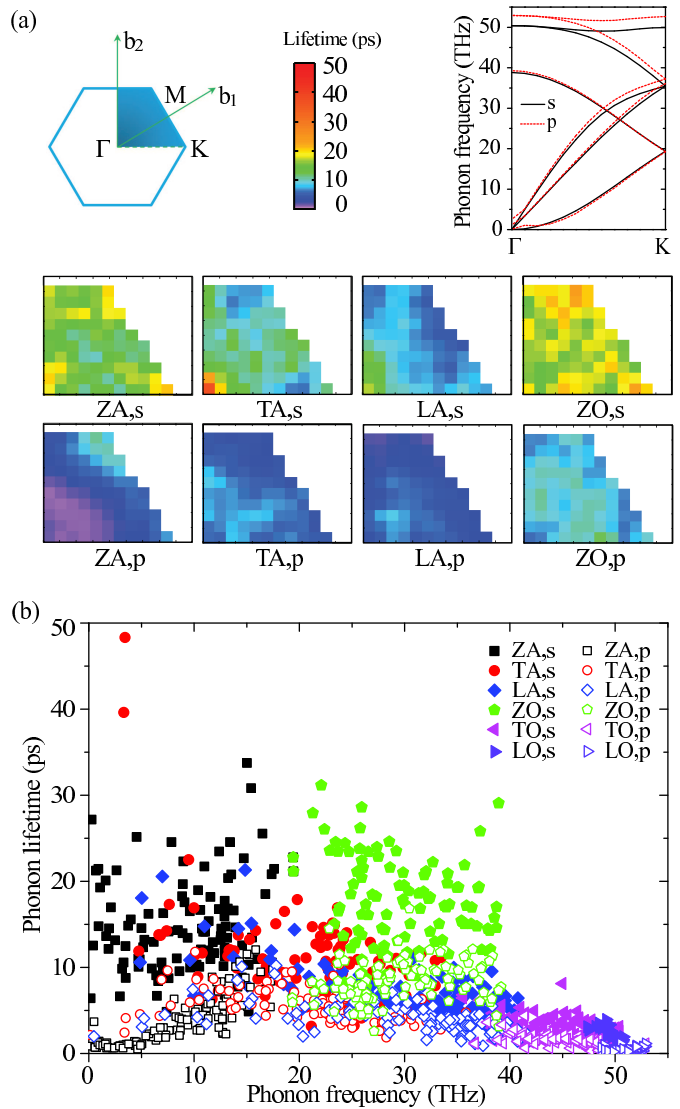}} \vspace{2mm}
\caption{(Color online) a) Distribution of phonon lifetime in the first quadrant of 1$^{st}$ BZ with the color representing lifetimes values. Also plotted are the phonon dispersions along $\Gamma$-K. b) Lifetime as a function of frequency for all phonon branches. The subscripts "s" and "p" are short for "suspended" and "supported", respectively.}
\label{fig:tau}
\end{figure}

The per-branch phonon lifetimes for both suspended and supported SLG at 304 K as a function of wavevector $\mathbf{k}$ and frequency are shown in Fig. \ref{fig:tau} (a) and (b), respectively. Because of symmetry, the phonon properties in the full BZ can be reproduced by only studying the irreducible $\mathbf{k}$-space, which is 1/12$^{th}$ of the 1$^{st}$ BZ. However, because of the simplicity in discretization and specifying group velocity along x direction, we study the 1$^{st}$ quadrant of BZ instead. From Fig. \ref{fig:tau} (a), we indeed find the distribution of lifetimes in suspended SLG to be reasonably symmetrical about high symmetry lines, considering finite discretization and uncertainty in predicted lifetime values. This confirms the self-consistency of our approach. In addition, as seen in Fig \ref{fig:tau} (a), the phonon dispersions in SLG with or without substrate do not differ much. Therefore, good correspondence between phonon modes in both cases can be well found to identify lifetime changes.

\subsubsection{Phonons in suspended graphene}

Existing studies either suggest negligible \cite{Nika_2009} or dominant \cite{Lindsay_2010} contribution from flexural phonons. A clear and correct picture would be crucial for understanding thermal transport in graphene. From Fig. \ref{fig:tau}, it is seen that, in suspended SLG, lifetime values of ZA phonons are in the range of sub 10-40 ps throughout BZ. The long lifetime of ZA phonons indicates relatively weak correlation between the in-plane and out-of-plane modes, which results in less scattering in ZA phonons \cite{Lindsay_2010}. It is interesting to see that out-of-plane optical (ZO) and ZA phonons have similarly large lifetimes, which can again be ascribed to the coupling strength between the in-plane and out-of-plane modes. The relatively weak dependence of ZA/ZO lifetime on the wavevector $\mathbf{k}$ is also an indication that their coupling to in-plane phonons are not very strong. As frequency increases, lifetime of ZA phonons first decreases then increases as $\mathbf{k}$ approaches BZ boundary. The relatively short lifetime of mid-frequency ZA phonons can be attributed to the fact that higher-order phonon scattering involving these modes is important and yields comparable scattering rates as that of three-phonon process, limiting the lifetime \cite{Ashcroft_1976}.

\begin{figure}
\centerline{\includegraphics{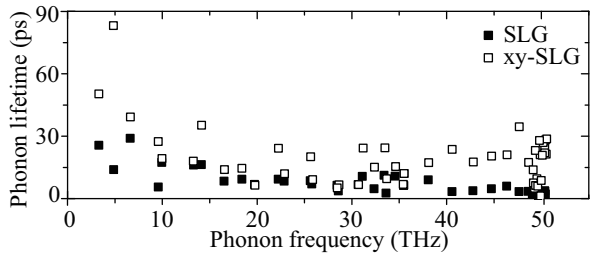}} \vspace{2mm}
\caption{In-plane phonon lifetime in xy-SLG as compared to that in suspended SLG along $\Gamma-K$ direction.}
\label{fig:3d2d}
\end{figure}

To further explore the role of flexural phonons in suspended SLG, we carry out simulations on suspended SLG with vibrations restricted to within the x-y plane (xy-SLG) so that the atomic motion is completely 2D. In doing this, out-of-plane phonons (ZA and ZO) are completely removed from the system. Predictions from GK method (not shown) consistently suggest several times larger $\kappa_l$ in xy-SLG. If only three-phonon scattering was considered, as suggested by Lindsay $\emph{et al}$, ZA phonons would contribute about 80$\%$ of the thermal conductivity \cite{Lindsay_2010}. Then the lifetime of remaining phonon modes should increase by at least four times to compensate the removal of phonon modes to result in an increased $\kappa_l$, according to Eq. (\ref{eqn:kl}). However, from SED analysis, as shown in the inset of Fig. \ref{fig:3d2d}, it is found that lifetime in xy-SLG is generally only two times longer than that in SLG in average. This implies the out-of-plane modes are coupled to in-plane modes with medium strength to leverage the scattering in in-plane and out-of-plane phonons in SLG. Such observation can only be understood when higher-order anharmonicities are taken into consideration. Due to the special symmetry of SLG, very stringent restrictions are imposed on the 3$^{rd}$ order scattering processes involving out-of-plane phonon modes \cite{Lindsay_2010}. There are so few 3$^{rd}$ order processes allowed that the higher-order scattering processes may yield considerable scattering rates and destroy the "pristine" transport of out-of-plane phonons \cite{Ashcroft_1976}, as supported by our simulation results.

Away from the zone center, the overall lifetime is mainly around 20 ps and below in most regions of BZ for longitudinal acoustic (LA) and transverse acoustic (TA) phonons. As frequency increases, lifetime shortens due to more efficient Umklapp scattering. For LA and TA phonons near zone center with small frequencies, similar to ZA phonons, their lifetime tends to diverge as seen in Fig. \ref{fig:tau} (b). This is because these phonon modes have very small wavevectors that, due to the momentum and energy conservation requirements, barely scatter with other phonons.

As a side note, it is seen that lifetime in SLG is actually not much higher than that in some solids with much lower thermal conductivity, such as solid argon \cite{Turney_2009}, silicon \cite{Henry_2008} and PbTe \cite{Qiu_2011_pbte}. Such observation indicates that despite the unique 2D transport in SLG, phonons do not necessarily experience much less scattering. The large $\kappa_l$ of SLG is mainly due to the small atomic weight of carbon atoms and the strong C-C bonds which lead to high phonon group velocity.

\subsubsection{Phonons in supported graphene}

When put on substrate, it is found that lifetimes are shortened throughout BZ and the full frequency range for all phonon branches. Namely, lifetimes of all acoustic phonons are generally shortened to be less than 10 ps. From the color map in Fig. \ref{fig:tau} (a), it can be clearly seen that such reduction is strongest for phonon modes closer to zone center. Also the distribution of lifetime in BZ becomes less symmetrical. This indicates the introduction of substrate strongly destructs the long-range order of phonon transport through random scattering sites at the interface. From Fig. \ref{fig:tau} (b) it is seen that the lifetime of low frequency phonons are mostly shortened. For frequency below 18 THz, the lifetime values monotonically decrease as frequency continuously reduces to zero, indicating the substrate couples most efficiently to low frequency acoustic phonons. One of the most significant outcomes of such observation is that the thermal conductivity of supported graphene will not be sensitive to graphene flake size. This is because long wavelength near-zone-center phonons in larger samples will have minimal lifetime due to substrate scattering and thus barely contribute to thermal transport.

Recent Brillouin scattering experiment on multilayer graphene supported on SiO$_2$/Si substrate \cite{Wang_2008} suggests the lifetime of near-zone-center LA/TA phonon is about 10-30 ps, which is shorter than that of suspended graphene, in line with our findings. In fact, due to the amorphous nature of silica, the coordinations of surface silicon/oxygen atoms cannot comply with the periodicity of SLG. As a result, the presence of the "irregularly" coordinated silicon/oxygen atoms at the interface will couple to the in-plane vibrations of carbon atoms as scattering centers and essentially hinder the in-plane phonons in SLG from transporting in a perfect 2D lattice. Therefore, to sustain long lifetime and thus high thermal conductivity of SLG, a substrate with better lattice match to SLG should be pursued. One model example is the multiple-layer graphene. When all but one layers are treated as substrate, the effective "SLG" is supported on a "substrate" with perfectly-matched lattice that the LA and TA phonons are less affected \cite{Ghosh_2010}. Therefore a supported double-layer graphene (DLG) or SLG supported on Boron Nitride (BN) substrate may turn out to be better for planar heat dissipation applications \cite{Wang_2011}.

The lifetime reduction is especially strong for ZA modes. This is due to the fact that the presence of the substrate breaks various symmetries (mirror-reflection, translational, etc.) in SLG and alters the out-of-plane vibrations of SLG by introducing SLG-substrate scattering which largely shortens the lifetime of ZA phonons. In this case, higher-order anharmonic interactions are no longer important. Similar arguments can also be made to explain the reduction of lifetime of ZO phonons.

\subsubsection{Longitudinal and transverse optical phonons}

The majority of longitudinal optical (LO) and transverse optical (TO) phonons have lifetimes about 1-4 ps at room temperature and reduced by about half when put on substrate. This is consistent with the optical phonon lifetimes of supported few-layer graphene measured by Raman spectroscopy \cite{Kang_2010, Wang_2010_ram, Calizo_2007}.

\subsection{Phonon mean free path}

\begin{figure}
\centerline{\includegraphics{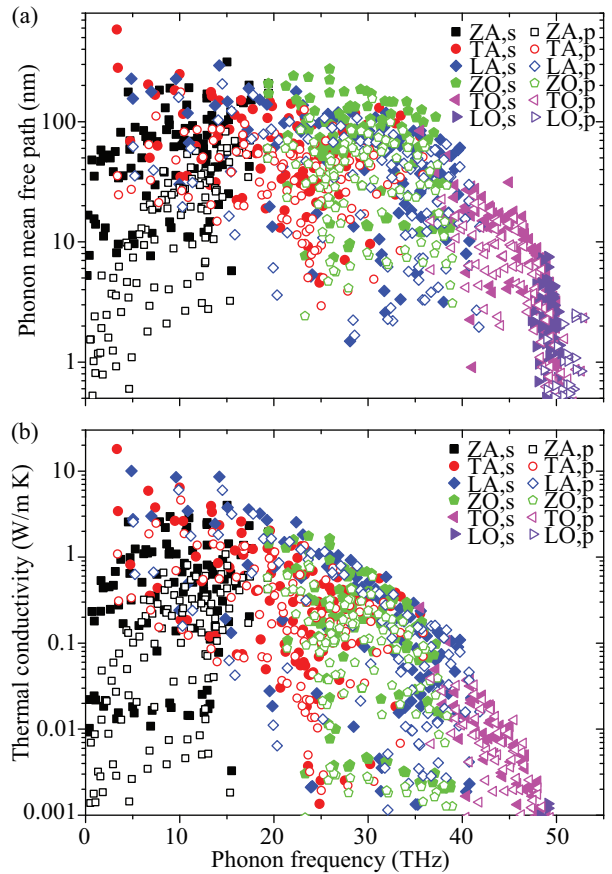}} \vspace{2mm}
\caption{(Color online) The frequency dependence of (a) MFP and (b) per-mode thermal conductivity for all phonon branches.}
\label{fig:mfpkl}
\end{figure}

As phonon spectral lifetimes are obtained for individual phonon modes in the preceding section, it is straightforward to compute phonon MFP based on its definition:
\begin{eqnarray}
l_{ph}=v_g \tau=\frac{d\omega}{dk}\tau.
\end{eqnarray}
As seen in Fig. \ref{fig:tau} (a), the phonon dispersion changes when SLG is put on substrate. This is due to the fact that the substrate induces internal stress and ripples in SLG, which leads to shift in phonon frequencies. The complete phonon dispersions before and after putting SLG on substrate are used to calculate the phonon group velocities, MFP's, thermal conductivities for suspended and supported SLG, respectively. MFP and per-mode thermal conductivity contribution of both suspended and supported SLG as a function of frequency are shown in Fig. \ref{fig:mfpkl}. It is seen that MFP ranges from 10 nm to 600 nm for all acoustic and ZO phonons while below 30 nm for LO and TO phonons.  MFP's of all acoustic and ZO phonons are generally longer than that of LO and TO phonons' due to their longer lifetime and higher group velocity. MFP's of LA and TA phonons exhibit the expected divergence when $\omega\rightarrow 0$ due to saturated group velocity and diverging lifetime. In contrast, no divergence is found for ZA modes because their group velocity $v_g\sim \sqrt{\omega}$ appears to decrease faster than the divergence of lifetime as $\omega\rightarrow 0$. When put on substrate, the MFP values drop to less than 200 nm for all acoustic and ZO phonons while those of LO and TO phonons are less affected. A comparison between the phonon dispersion relations before and after the presence of substrate doesn't suggest significant change in phonon group velocities, as seen in Fig. \ref{fig:tau} (a). Therefore, according to Eq. (\ref{eqn:kl}), the effect of substrate in limiting SLG thermal conductivity should be mostly attributed to the introduction of SLG-substrate scattering that shortens lifetime. Due to the small MFP in supported SLG, again, we do not expect graphene flake size to significantly affect the thermal conductivity as long as it is wider than 200 nm.

\subsection{Thermal conductivity decomposition}

From Fig. \ref{fig:mfpkl} (b), it is seen that the majority of thermal conductivity contributions are from phonons with lower frequencies. This can be understood from Eq. (\ref{eqn:kl}) that $v_g$ and $\tau$ are large for lower frequencies and $c_v$ decreases with increasing frequency due to the change in phonon population. Therefore, despite their small group velocity, ZA phonons still contribute about $1/4$ of the thermal conductivity, which is not much less than that from LA and TA phonons. Such contribution is mainly from mid-frequency phonons due to the different frequency dependence of ZA group velocity and lifetime. When put on substrate, the thermal conductivity contributed from all acoustic and ZO phonon branches are reduced, owing to the shortened lifetime due to SLG-substrate scattering mentioned in the above text.

\begin{figure}
\centerline{\includegraphics{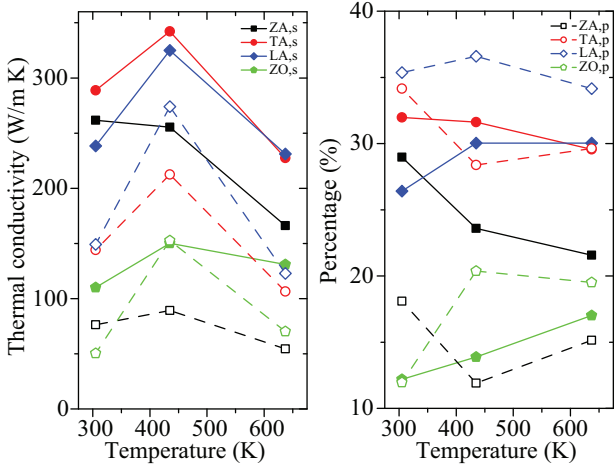}} \vspace{2mm}
\caption{(Color online) a) Contributions to thermal conductivity from ZA, TA, LA and ZO phonons. b) Percentage of contributions to thermal conductivity from ZA, TA, LA and ZO phonons.}
\label{fig:kldecomp}
\end{figure}

The relative importance of contribution to thermal conductivity from all phonon branches (except TO and LO) at elevated temperatures are computed using Eq. (\ref{eqn:kl}) and presented in Fig. \ref{fig:kldecomp}. It is seen that all acoustic phonons contribute about 20-35$\%$ to the total thermal conductivity in suspended SLG. The relative contribution from ZA phonons increases with decreasing temperature as other phonons freeze out. When put on substrate, contribution from all branches to the total thermal conductivity reduces. It can also be seen that the reduction in ZA phonon contributions are most severe as its relative contribution drops to be around 15$\%$, comparable to that of ZO phonons. In contrast, the relative contributions from both LA and TA phonons are still above 30$\%$ when supported. Therefore, the in-plane acoustic phonons are mainly responsible for the thermal transport in supported SLG, consistent with the theory of Seol \emph{et al} \cite{Seol_2010}. At higher temperature, $\kappa_l$ values associated with all acoustic phonons decrease because of stronger Umklapp scattering. At lower temperature, $\kappa_l$ values increase owing to the activation of higher frequency phonon modes, which is only captured after QC.

\begin{figure}
\centerline{\includegraphics{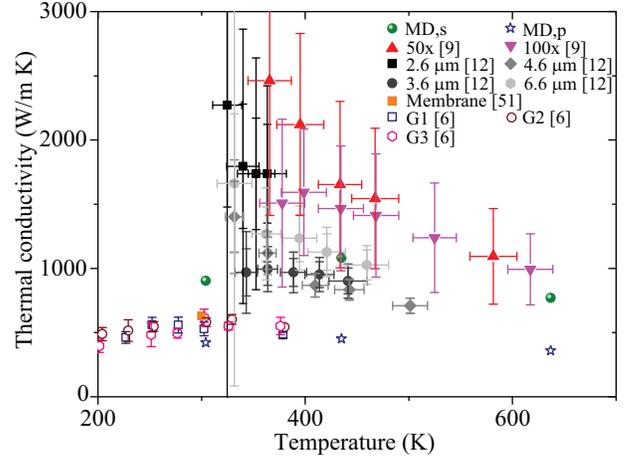}} \vspace{2mm}
\caption{(Color online) Predicted total thermal conductivity values in comparison with experiments of suspended SLG \cite{Cai_2010, Lee_2011}, membrane \cite{Faugeras_2010} and SLG supported on SiO$_2$ substrate \cite{Seol_2010}.}
\label{fig:kl}
\end{figure}

The predicted total thermal conductivity for suspended graphene before QC is 1626 W/m K at 300 K, which agrees with $\kappa_l=1695$ W/m K obtained from Green-Kubo method. The experimentally measured thermal conductivity values for suspended SLG spread widely, which can be partly attributed to the varied graphene sample quality. As mentioned before, because of the large uncertainty and spread data, no clear temperature trend or graphene flake size dependence can be demonstrated experimentally so far. Nonetheless, as seen in Fig. \ref{fig:kl}, the $\kappa_l$ values predicted from MD simulations are consistent with experiments for both suspended SLG and SLG on amorphous SiO$_2$ substrate. Such consistency and the successful prediction of thermal conductivity reduction in supported SLG without adjustable parameters suggest the validity of the methodology and the results presented in this work.

\section{Further explorations}

\subsection{Effects of interfacial bonding}

\begin{figure}
\centerline{\includegraphics{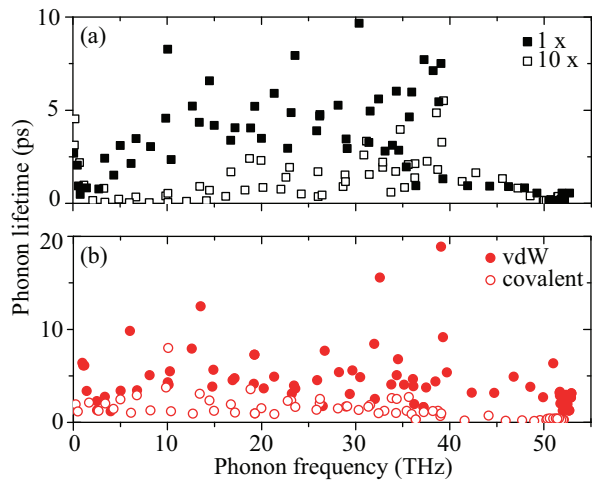}} \vspace{2mm}
\caption{(Color online) Lifetime of all phonon branches along $\Gamma-K$ in supported SLG. a) SiO$_2$ substrate, when 10 times stronger LJ interactions assumed. b) Silicon substrate, when covalent interfacial interactions are assumed.}
\label{fig:vdwcov}
\end{figure}

The above results for supported SLG are obtained under the assumption that the interfacial interactions between SLG and substrate is of vdW type \cite{Miwa_2005}. On the other hand, it is found that covalent bonds can form between graphene and SiC substrate \cite{Mattausch_2007}. To explore the effects of interfacial bondings on the thermal transport in supported SLG, we study a) lifetimes when LJ interactions are assumed to be 10 times stronger than used in previous sections; b) lifetimes of graphene on silicon substrate with either vdW interactions or covalent bonds at the interface. For the latter case, we use LJ potentials with parameters $\epsilon_{C-Si}=8.909$ meV, $\sigma_{C-Si}=3.326$ $\AA$ to model vdW interactions while using OPT with mixing rules to model the C-Si covalent bonds. SED analysis is performed and the results are compared in $\Gamma-K$ direction, as shown in Fig. \ref{fig:vdwcov} (a) and (b). As seen, when the vdW interactions are 10 times stronger or all bonds at the interface are covalent, the phonon scattering in SLG due to substrate is much stronger, leading to significantly shortened lifetime and thus strongly reduced thermal conductivity. According to the studies on charge density distribution at CNT/silicon interface \cite{Orellana_2008}, the interactions between SLG and a substrate with non-passivated surface appear to be a mixture of non-bonded vdW interactions and covalent bonds. Therefore, in general, the value of thermal conductivity of supported SLG may be a weighted average of the values when both vdW and covalent bonds at the interface are assumed.

On the other hand, we also find similar reduction in thermal conductivity when silicon substrate is used in the simulation with vdW interaction strength similar to that assumed in the preceding sections where SiO$_2$ substrate is used \cite{Qiu_2011_mrs}. Therefore, we hypothesize that the interaction strength and geometry at SLG-substrate interface should be deterministic factors for thermal conductivity reduction in supported SLG. Our theory is supported by recent experiment that polymeric residue can bring the thermal conductivity of suspended DLG down to around 600 W/m K \cite{Pettes_2011}, which is similar to that found in SLG supported on SiO$_2$ substrate \cite{Seol_2010}.

\subsection{Effects of interatomic potentials}

It is known that REBO potentials \cite{Brenner_2002} for hydrocarbons tend to underestimate the thermal conductivity of graphene nanoribbons (GNR) and CNT's, possibly due to the inadequateness of these potentials in representing the anharmonicity in carbon-based systems. Nonetheless, due to the popularity of REBO potentials in literature, we also use them to perform MD simulations as well as SED analysis for both suspended and supported SLG then compare to the results from OPT, as summarized in Table \ref{tab:rebo}. As seen, despite the underestimation in total thermal conductivity values, REBO potentials can similarly reproduce relative importance of phonon branches in contributing to $\kappa_l$ and the reduction of $\kappa_l$ when SLG is put on substrate. However, it does fail to predict significant reduction in ZA contribution in supported SLG. Therefore, the use of REBO potentials to study carbon-based systems should generally be valid in a qualitative sense while caution is advised.

\begin{table}
\caption{\label{tab:rebo}Summary of thermal conductivity and its decomposition in suspended and supported SLG predicted by OPT and REBO potentials at 435 K. The units for $\kappa_l$ and individual mode contributions are (W/m K) and ($\%$), respectively.}
\begin{tabular}{l|ccccr}
\hline
Suspended & $\kappa_l$ & ZA & TA & LA & ZO \\
\hline
OPT & 1082 & 23.6 & 31.6 & 30.0 & 13.9 \\
\hline
REBO & 509 & 33.8 & 27.7 & 30.5 & 7.9 \\
\hline
Supported  & $\kappa_l$ & ZA & TA & LA & ZO \\
\hline
OPT & 451 & 16.8 & 32.5 & 33.8 & 16.0 \\
\hline
REBO & 224 & 27.5 & 28.7 & 29.5 & 13.7 \\
\hline
\end{tabular}
\end{table}

\section{Conclusions}

In conclusion, we predict the phonon lifetime, MFP and thermal conductivity of both suspended and supported SLG using MD simulations and SED analysis. Our results are found to be consistent with both optical and thermal measurements for phonon lifetime and thermal conductivity of SLG in literature. In suspended SLG, out-of-plane phonons (ZA and ZO) are found to couple to the in-plane phonons in the Umklapp scattering due to both 3$^{rd}$ and higher-order anharmonic interactions, leading to around 25-30$\%$ contribution to thermal conductivity from ZA phonons. In the presence of substrate, the contributions to thermal conductivity from all acoustic and ZO phonon branches are reduced, owing to the SLG-substrate scattering and the breakdown of symmetries in both in-plane and out-of-plane phonons. ZA phonon contributions to thermal conductivity are largely suppressed to be around 15$\%$ in supported SLG, making LA and TA phonons most responsible for thermal transport when SLG is put on amorphous SiO$_2$ substrate. We suggest to use substrates with closer lattice match to SLG for better planar heat dissipations.


\begin{acknowledgments}
This work is supported by the Air Force Office of Scientific Research (AFOSR) through the Discovery Challenge Thrust (DCT) Program (grant FA9550-11-1-0057, Program Manager Kumar Jata) and by the Cooling Technologies Research Center (CTRC), an NSF University/Industry Cooperation Research Center. We thank Alexander A. Balandin, Alan J. H. McGaughey, Jayathi Murthy, Asegun Henry, Joseph E. Turney and Dhruv Singh for helpful discussions.
\end{acknowledgments}

\appendix

\section{Quantum Corrections}

In MD, quantum effects in specific heat and phonon occupation number are not captured due to its classical nature. Therefore, MD simulations are generally not readily suitable to study thermal transport at temperatures much lower than the Debye temperature. While it is difficult to explicitly include quantum effects in MD simulations \cite{Turney_2009_qc}, there are some quantum correction (QC) approaches available based on $\emph{ad hoc}$ physical arguments. According to them, the temperature correction is made by equating the total energies of the classical and quantum systems:
\begin{equation}
E_C(T_C)=E_Q(T_Q),
\end{equation}
and the thermal conductivity correction is obtained as:
\begin{equation}
\kappa_Q=\kappa_C\frac{dT_C}{dT_Q}.
\end{equation}
While overall $\kappa$ values from such QCs may seem okey, the correctness of individual phonon mode properties can not be guaranteed \cite{Turney_2009_qc}. Fortunately, we are able to obtain individual phonon properties from the spectral energy density (SED) analysis, enabling QCs in a mode-dependent manner. The phonon lifetime is largely determined by the phonon distribution in the system. Therefore, to maximally ensure the validness of MD simulation in reproducing the mode-dependent phonon scattering, we perform QC by finding a quantum system at temperature $T_Q$ that has the closest phonon distribution to the classical system at temperature $T_C$. That is to find a $T_Q$ that leads to the minimum of function
\begin{equation}
M(T_Q,T_C)=\int_0^{\omega_{max}}|f_Q(T_Q,\omega)-f_C(T_C,\omega)| D(\omega) d\omega
\end{equation}
for a choice of $T_C$. Here $f_Q(T_Q,\omega)=1/[\exp(x_Q)-1]$ and $f_C(T_C,\omega)=1/x_C$ are the quantum and classical phonon occupation numbers, respectively, where $x_i=\hbar\omega/k_B T_i$. $D(\omega)$ is the phonon density of states that can be directly extracted from MD simulations through velocity-velocity autocorrelations. As an example, for suspended single-layer graphene (SLG) at $T_C=190$ K, the equivalent quantum system is found to be at $T_Q=304$ K.

Once $T_Q$ is found, we have $\tau_Q(\omega_{\mathbf{k},\nu}, T_Q)\approx \tau_C(\omega_{\mathbf{k},\nu}, T_C)$. By taking into account the quantum specific heat
\begin{equation}
c_v^Q=\frac{k_B x_Q^2 \exp(x_Q)}{V[\exp(x_Q)-1]^2},
\end{equation}
a correction to thermal conductivity can be obtained as:
\begin{equation}
\kappa_{\alpha,Q}=\sum_{\mathbf{k}}\sum_{\nu}c_v^Q v_{g,\alpha}^2 \tau_Q.
\end{equation}


\begin{thebibliography}{55}
\expandafter\ifx\csname natexlab\endcsname\relax\def\natexlab#1{#1}\fi
\expandafter\ifx\csname bibnamefont\endcsname\relax
  \def\bibnamefont#1{#1}\fi
\expandafter\ifx\csname bibfnamefont\endcsname\relax
  \def\bibfnamefont#1{#1}\fi
\expandafter\ifx\csname citenamefont\endcsname\relax
  \def\citenamefont#1{#1}\fi
\expandafter\ifx\csname url\endcsname\relax
  \def\url#1{\texttt{#1}}\fi
\expandafter\ifx\csname urlprefix\endcsname\relax\def\urlprefix{URL }\fi
\providecommand{\bibinfo}[2]{#2}
\providecommand{\eprint}[2][]{\url{#2}}

\bibitem[{\citenamefont{Novoselov et~al.}(2004)\citenamefont{Novoselov, Geim,
  Morozov, Jiang, Zhang, Dubonos, Grigorieva1, and Firsov}}]{Novoselov_2004}
\bibinfo{author}{\bibfnamefont{K.~S.} \bibnamefont{Novoselov}},
  \bibinfo{author}{\bibfnamefont{A.~K.} \bibnamefont{Geim}},
  \bibinfo{author}{\bibfnamefont{S.~V.} \bibnamefont{Morozov}},
  \bibinfo{author}{\bibfnamefont{D.}~\bibnamefont{Jiang}},
  \bibinfo{author}{\bibfnamefont{Y.}~\bibnamefont{Zhang}},
  \bibinfo{author}{\bibfnamefont{S.~V.} \bibnamefont{Dubonos}},
  \bibinfo{author}{\bibfnamefont{I.~V.} \bibnamefont{Grigorieva1}},
  \bibnamefont{and} \bibinfo{author}{\bibfnamefont{A.~A.}
  \bibnamefont{Firsov}}, \bibinfo{journal}{Science}
  \textbf{\bibinfo{volume}{306}}, \bibinfo{pages}{666} (\bibinfo{year}{2004}).

\bibitem[{\citenamefont{Bolotin et~al.}(2008)\citenamefont{Bolotin, Sikes,
  Jiang, Klim, Fudenberg, Hone, Kim, and Stormer}}]{Bolotin_2008}
\bibinfo{author}{\bibfnamefont{K.~I.} \bibnamefont{Bolotin}},
  \bibinfo{author}{\bibfnamefont{K.~J.} \bibnamefont{Sikes}},
  \bibinfo{author}{\bibfnamefont{Z.}~\bibnamefont{Jiang}},
  \bibinfo{author}{\bibfnamefont{M.}~\bibnamefont{Klim}},
  \bibinfo{author}{\bibfnamefont{G.}~\bibnamefont{Fudenberg}},
  \bibinfo{author}{\bibfnamefont{J.}~\bibnamefont{Hone}},
  \bibinfo{author}{\bibfnamefont{P.}~\bibnamefont{Kim}}, \bibnamefont{and}
  \bibinfo{author}{\bibfnamefont{H.~L.} \bibnamefont{Stormer}},
  \bibinfo{journal}{Solid State Comm.} \textbf{\bibinfo{volume}{146}},
  \bibinfo{pages}{351} (\bibinfo{year}{2008}).

\bibitem[{\citenamefont{Lee et~al.}(2008)\citenamefont{Lee, Wei, Kysar, and
  Hone}}]{Lee_2008}
\bibinfo{author}{\bibfnamefont{C.}~\bibnamefont{Lee}},
  \bibinfo{author}{\bibfnamefont{X.}~\bibnamefont{Wei}},
  \bibinfo{author}{\bibfnamefont{J.~W.} \bibnamefont{Kysar}}, \bibnamefont{and}
  \bibinfo{author}{\bibfnamefont{J.}~\bibnamefont{Hone}},
  \bibinfo{journal}{Science} \textbf{\bibinfo{volume}{321}},
  \bibinfo{pages}{385} (\bibinfo{year}{2008}).

\bibitem[{\citenamefont{Balandin et~al.}(2008)\citenamefont{Balandin, Ghosh,
  Bao, Calizo, Teweldebrhan, Miao, and Lau}}]{Balandin_2008}
\bibinfo{author}{\bibfnamefont{A.~A.} \bibnamefont{Balandin}},
  \bibinfo{author}{\bibfnamefont{S.}~\bibnamefont{Ghosh}},
  \bibinfo{author}{\bibfnamefont{W.}~\bibnamefont{Bao}},
  \bibinfo{author}{\bibfnamefont{I.}~\bibnamefont{Calizo}},
  \bibinfo{author}{\bibfnamefont{D.}~\bibnamefont{Teweldebrhan}},
  \bibinfo{author}{\bibfnamefont{F.}~\bibnamefont{Miao}}, \bibnamefont{and}
  \bibinfo{author}{\bibfnamefont{C.~N.} \bibnamefont{Lau}},
  \bibinfo{journal}{Nano Lett.} \textbf{\bibinfo{volume}{8}},
  \bibinfo{pages}{902} (\bibinfo{year}{2008}).

\bibitem[{\citenamefont{Geim and Novoselov}(2007)}]{Geim_2007}
\bibinfo{author}{\bibfnamefont{A.~K.} \bibnamefont{Geim}} \bibnamefont{and}
  \bibinfo{author}{\bibfnamefont{K.~S.} \bibnamefont{Novoselov}},
  \bibinfo{journal}{Nature Mater.} \textbf{\bibinfo{volume}{6}},
  \bibinfo{pages}{183} (\bibinfo{year}{2007}).

\bibitem[{\citenamefont{Seol et~al.}(2010)\citenamefont{Seol, Jo, Moore,
  Lindsay, Aitken, Pettes, Li, Yao, Huang, Broido et~al.}}]{Seol_2010}
\bibinfo{author}{\bibfnamefont{J.~H.} \bibnamefont{Seol}},
  \bibinfo{author}{\bibfnamefont{I.}~\bibnamefont{Jo}},
  \bibinfo{author}{\bibfnamefont{A.~L.} \bibnamefont{Moore}},
  \bibinfo{author}{\bibfnamefont{L.}~\bibnamefont{Lindsay}},
  \bibinfo{author}{\bibfnamefont{Z.~H.} \bibnamefont{Aitken}},
  \bibinfo{author}{\bibfnamefont{M.~T.} \bibnamefont{Pettes}},
  \bibinfo{author}{\bibfnamefont{X.}~\bibnamefont{Li}},
  \bibinfo{author}{\bibfnamefont{Z.}~\bibnamefont{Yao}},
  \bibinfo{author}{\bibfnamefont{R.}~\bibnamefont{Huang}},
  \bibinfo{author}{\bibfnamefont{D.}~\bibnamefont{Broido}},
  \bibnamefont{et~al.}, \bibinfo{journal}{Science}
  \textbf{\bibinfo{volume}{328}}, \bibinfo{pages}{213} (\bibinfo{year}{2010}).

\bibitem[{\citenamefont{Balandin}(2011)}]{Balandin_2011}
\bibinfo{author}{\bibfnamefont{A.~A.} \bibnamefont{Balandin}},
  \bibinfo{journal}{Nature Mater.} \textbf{\bibinfo{volume}{10}},
  \bibinfo{pages}{569} (\bibinfo{year}{2011}).

\bibitem[{\citenamefont{Basile et~al.}(2006)\citenamefont{Basile, Bernardin,
  and Olla}}]{Basile_2006}
\bibinfo{author}{\bibfnamefont{G.}~\bibnamefont{Basile}},
  \bibinfo{author}{\bibfnamefont{C.}~\bibnamefont{Bernardin}},
  \bibnamefont{and} \bibinfo{author}{\bibfnamefont{S.}~\bibnamefont{Olla}},
  \bibinfo{journal}{Phys. Rev. Lett.} \textbf{\bibinfo{volume}{96}},
  \bibinfo{pages}{204303} (\bibinfo{year}{2006}).

\bibitem[{\citenamefont{Cai et~al.}(2010)\citenamefont{Cai, Moore, Zhu, Li,
  Chen, Shi, and Ruoff}}]{Cai_2010}
\bibinfo{author}{\bibfnamefont{W.}~\bibnamefont{Cai}},
  \bibinfo{author}{\bibfnamefont{A.~L.} \bibnamefont{Moore}},
  \bibinfo{author}{\bibfnamefont{Y.}~\bibnamefont{Zhu}},
  \bibinfo{author}{\bibfnamefont{X.}~\bibnamefont{Li}},
  \bibinfo{author}{\bibfnamefont{S.}~\bibnamefont{Chen}},
  \bibinfo{author}{\bibfnamefont{L.}~\bibnamefont{Shi}}, \bibnamefont{and}
  \bibinfo{author}{\bibfnamefont{R.~S.} \bibnamefont{Ruoff}},
  \bibinfo{journal}{Nano Lett.} \textbf{\bibinfo{volume}{10}},
  \bibinfo{pages}{1645} (\bibinfo{year}{2010}).

\bibitem[{\citenamefont{Ghosh et~al.}(2010)\citenamefont{Ghosh, Bao, Nika,
  Subrina, Pokatilov, Lau, and Balandin}}]{Ghosh_2010}
\bibinfo{author}{\bibfnamefont{S.}~\bibnamefont{Ghosh}},
  \bibinfo{author}{\bibfnamefont{W.}~\bibnamefont{Bao}},
  \bibinfo{author}{\bibfnamefont{D.~L.} \bibnamefont{Nika}},
  \bibinfo{author}{\bibfnamefont{S.}~\bibnamefont{Subrina}},
  \bibinfo{author}{\bibfnamefont{E.~P.} \bibnamefont{Pokatilov}},
  \bibinfo{author}{\bibfnamefont{C.~N.} \bibnamefont{Lau}}, \bibnamefont{and}
  \bibinfo{author}{\bibfnamefont{A.~A.} \bibnamefont{Balandin}},
  \bibinfo{journal}{Nature Materials} \textbf{\bibinfo{volume}{9}},
  \bibinfo{pages}{555} (\bibinfo{year}{2010}).

\bibitem[{\citenamefont{Qiu and Ruan}(2010)}]{Qiu_2010}
\bibinfo{author}{\bibfnamefont{B.}~\bibnamefont{Qiu}} \bibnamefont{and}
  \bibinfo{author}{\bibfnamefont{X.}~\bibnamefont{Ruan}},
  \bibinfo{journal}{Appl. Phys. Lett.} \textbf{\bibinfo{volume}{97}},
  \bibinfo{pages}{183107} (\bibinfo{year}{2010}).

\bibitem[{\citenamefont{Lee et~al.}(2011)\citenamefont{Lee, Yoon, Kim, Lee, and
  Cheong}}]{Lee_2011}
\bibinfo{author}{\bibfnamefont{J.-U.} \bibnamefont{Lee}},
  \bibinfo{author}{\bibfnamefont{D.}~\bibnamefont{Yoon}},
  \bibinfo{author}{\bibfnamefont{H.}~\bibnamefont{Kim}},
  \bibinfo{author}{\bibfnamefont{S.~W.} \bibnamefont{Lee}}, \bibnamefont{and}
  \bibinfo{author}{\bibfnamefont{H.}~\bibnamefont{Cheong}},
  \bibinfo{journal}{Phys. Rev. B} \textbf{\bibinfo{volume}{83}},
  \bibinfo{pages}{081419(R)} (\bibinfo{year}{2011}).

\bibitem[{\citenamefont{Hu et~al.}(2009)\citenamefont{Hu, Ruan, and
  Chen}}]{Hu_2009}
\bibinfo{author}{\bibfnamefont{J.}~\bibnamefont{Hu}},
  \bibinfo{author}{\bibfnamefont{X.}~\bibnamefont{Ruan}}, \bibnamefont{and}
  \bibinfo{author}{\bibfnamefont{Y.~P.} \bibnamefont{Chen}},
  \bibinfo{journal}{Nano Lett.} \textbf{\bibinfo{volume}{9}},
  \bibinfo{pages}{2730} (\bibinfo{year}{2009}).

\bibitem[{\citenamefont{Guo et~al.}(2009)\citenamefont{Guo, Zhang, and
  Gong}}]{Guo_2009}
\bibinfo{author}{\bibfnamefont{Z.}~\bibnamefont{Guo}},
  \bibinfo{author}{\bibfnamefont{D.}~\bibnamefont{Zhang}}, \bibnamefont{and}
  \bibinfo{author}{\bibfnamefont{X.~G.} \bibnamefont{Gong}},
  \bibinfo{journal}{Appl. Phys. Lett.} \textbf{\bibinfo{volume}{95}},
  \bibinfo{pages}{163103} (\bibinfo{year}{2009}).

\bibitem[{\citenamefont{Lindsay et~al.}(2010)\citenamefont{Lindsay, Broido, and
  Mingo}}]{Lindsay_2010}
\bibinfo{author}{\bibfnamefont{L.}~\bibnamefont{Lindsay}},
  \bibinfo{author}{\bibfnamefont{D.~A.} \bibnamefont{Broido}},
  \bibnamefont{and} \bibinfo{author}{\bibfnamefont{N.}~\bibnamefont{Mingo}},
  \bibinfo{journal}{Phys. Rev. B} \textbf{\bibinfo{volume}{82}},
  \bibinfo{pages}{115427} (\bibinfo{year}{2010}).

\bibitem[{\citenamefont{Tan et~al.}(2011)\citenamefont{Tan, Wang, and
  Gan}}]{Tan_2011}
\bibinfo{author}{\bibfnamefont{Z.~W.} \bibnamefont{Tan}},
  \bibinfo{author}{\bibfnamefont{J.-S.} \bibnamefont{Wang}}, \bibnamefont{and}
  \bibinfo{author}{\bibfnamefont{C.~K.} \bibnamefont{Gan}},
  \bibinfo{journal}{Nano Lett.} \textbf{\bibinfo{volume}{11}},
  \bibinfo{pages}{214} (\bibinfo{year}{2011}).

\bibitem[{\citenamefont{Nika et~al.}(2009)\citenamefont{Nika, Pokatilov,
  Askerov, and Balandin}}]{Nika_2009}
\bibinfo{author}{\bibfnamefont{D.~L.} \bibnamefont{Nika}},
  \bibinfo{author}{\bibfnamefont{E.~P.} \bibnamefont{Pokatilov}},
  \bibinfo{author}{\bibfnamefont{A.~S.} \bibnamefont{Askerov}},
  \bibnamefont{and} \bibinfo{author}{\bibfnamefont{A.~A.}
  \bibnamefont{Balandin}}, \bibinfo{journal}{Phys. Rev. B}
  \textbf{\bibinfo{volume}{79}}, \bibinfo{pages}{155413}
  (\bibinfo{year}{2009}).

\bibitem[{\citenamefont{Ong and Pop}(2011)}]{Ong_2011}
\bibinfo{author}{\bibfnamefont{Z.-Y.} \bibnamefont{Ong}} \bibnamefont{and}
  \bibinfo{author}{\bibfnamefont{E.}~\bibnamefont{Pop}},
  \bibinfo{journal}{Phys. Rev. B} \textbf{\bibinfo{volume}{84}},
  \bibinfo{pages}{075471} (\bibinfo{year}{2011}).

\bibitem[{\citenamefont{Ong and Pop}(2010)}]{Ong_2010}
\bibinfo{author}{\bibfnamefont{Z.-Y.} \bibnamefont{Ong}} \bibnamefont{and}
  \bibinfo{author}{\bibfnamefont{E.}~\bibnamefont{Pop}},
  \bibinfo{journal}{Phys. Rev. B} \textbf{\bibinfo{volume}{81}},
  \bibinfo{pages}{155408} (\bibinfo{year}{2010}).

\bibitem[{\citenamefont{Miwa et~al.}(2005)\citenamefont{Miwa, Orellana, and
  Fazzio}}]{Miwa_2005}
\bibinfo{author}{\bibfnamefont{R.~H.} \bibnamefont{Miwa}},
  \bibinfo{author}{\bibfnamefont{W.}~\bibnamefont{Orellana}}, \bibnamefont{and}
  \bibinfo{author}{\bibfnamefont{A.}~\bibnamefont{Fazzio}},
  \bibinfo{journal}{Applied Surface Science} \textbf{\bibinfo{volume}{244}},
  \bibinfo{pages}{124} (\bibinfo{year}{2005}).

\bibitem[{\citenamefont{Vallabhaneni et~al.}(2011)\citenamefont{Vallabhaneni,
  Qiu, Hu, Chen, and Ruan}}]{Vallabhaneni_2011}
\bibinfo{author}{\bibfnamefont{A.~K.} \bibnamefont{Vallabhaneni}},
  \bibinfo{author}{\bibfnamefont{B.}~\bibnamefont{Qiu}},
  \bibinfo{author}{\bibfnamefont{J.}~\bibnamefont{Hu}},
  \bibinfo{author}{\bibfnamefont{Y.~P.} \bibnamefont{Chen}}, \bibnamefont{and}
  \bibinfo{author}{\bibfnamefont{X.}~\bibnamefont{Ruan}}, \bibinfo{journal}{in
  preparation}  (\bibinfo{year}{2011}).

\bibitem[{\citenamefont{Orellana}(2008)}]{Orellana_2008}
\bibinfo{author}{\bibfnamefont{W.}~\bibnamefont{Orellana}},
  \bibinfo{journal}{Appl. Phys. Lett.} \textbf{\bibinfo{volume}{92}},
  \bibinfo{pages}{093109} (\bibinfo{year}{2008}).

\bibitem[{\citenamefont{Plimpton}(1995)}]{Plimpton_1995}
\bibinfo{author}{\bibfnamefont{S.}~\bibnamefont{Plimpton}},
  \bibinfo{journal}{J. Comp. Phys.} \textbf{\bibinfo{volume}{117}},
  \bibinfo{pages}{1} (\bibinfo{year}{1995}).

\bibitem[{\citenamefont{Lindsay and Broido}(2010)}]{Lindsay_2010_1}
\bibinfo{author}{\bibfnamefont{L.}~\bibnamefont{Lindsay}} \bibnamefont{and}
  \bibinfo{author}{\bibfnamefont{D.~A.} \bibnamefont{Broido}},
  \bibinfo{journal}{Phys. Rev. B} \textbf{\bibinfo{volume}{81}},
  \bibinfo{pages}{205441} (\bibinfo{year}{2010}).

\bibitem[{\citenamefont{Munetoh et~al.}(2007)\citenamefont{Munetoh, Motooka,
  Moriguchi, and Shintani}}]{Munetoh_2007}
\bibinfo{author}{\bibfnamefont{S.}~\bibnamefont{Munetoh}},
  \bibinfo{author}{\bibfnamefont{T.}~\bibnamefont{Motooka}},
  \bibinfo{author}{\bibfnamefont{K.}~\bibnamefont{Moriguchi}},
  \bibnamefont{and} \bibinfo{author}{\bibfnamefont{A.}~\bibnamefont{Shintani}},
  \bibinfo{journal}{Comput. Mater. Sci.} \textbf{\bibinfo{volume}{39}},
  \bibinfo{pages}{334} (\bibinfo{year}{2007}).

\bibitem[{\citenamefont{Rappe et~al.}(1992)\citenamefont{Rappe, Casewit,
  Colwell, Goddard, and Skiff}}]{Rappe_1992}
\bibinfo{author}{\bibfnamefont{A.~K.} \bibnamefont{Rappe}},
  \bibinfo{author}{\bibfnamefont{C.~J.} \bibnamefont{Casewit}},
  \bibinfo{author}{\bibfnamefont{K.~S.} \bibnamefont{Colwell}},
  \bibinfo{author}{\bibfnamefont{W.~A.} \bibnamefont{Goddard}},
  \bibnamefont{and} \bibinfo{author}{\bibfnamefont{W.~M.} \bibnamefont{Skiff}},
  \bibinfo{journal}{J. Am. Chem. Soc.} \textbf{\bibinfo{volume}{114}},
  \bibinfo{pages}{10024} (\bibinfo{year}{1992}).

\bibitem[{\citenamefont{Tewary and Yang}(2009)}]{Tewary_2009}
\bibinfo{author}{\bibfnamefont{V.~K.} \bibnamefont{Tewary}} \bibnamefont{and}
  \bibinfo{author}{\bibfnamefont{B.}~\bibnamefont{Yang}},
  \bibinfo{journal}{Phys. Rev. B} \textbf{\bibinfo{volume}{79}},
  \bibinfo{pages}{125416} (\bibinfo{year}{2009}).

\bibitem[{\citenamefont{Che et~al.}(2000)\citenamefont{Che, Cagin, Deng, and
  III}}]{Che_2000}
\bibinfo{author}{\bibfnamefont{J.}~\bibnamefont{Che}},
  \bibinfo{author}{\bibfnamefont{T.}~\bibnamefont{Cagin}},
  \bibinfo{author}{\bibfnamefont{W.}~\bibnamefont{Deng}}, \bibnamefont{and}
  \bibinfo{author}{\bibfnamefont{W.~A.~G.} \bibnamefont{III}},
  \bibinfo{journal}{Journal of Chemical Physics}
  \textbf{\bibinfo{volume}{113}}, \bibinfo{pages}{6888} (\bibinfo{year}{2000}).

\bibitem[{\citenamefont{Thomas et~al.}(2010)\citenamefont{Thomas, Turney,
  Iutzi, Amon, and McGaughey}}]{Thomas_2010}
\bibinfo{author}{\bibfnamefont{J.~A.} \bibnamefont{Thomas}},
  \bibinfo{author}{\bibfnamefont{J.~E.} \bibnamefont{Turney}},
  \bibinfo{author}{\bibfnamefont{R.~M.} \bibnamefont{Iutzi}},
  \bibinfo{author}{\bibfnamefont{C.~H.} \bibnamefont{Amon}}, \bibnamefont{and}
  \bibinfo{author}{\bibfnamefont{A.~J.~H.} \bibnamefont{McGaughey}},
  \bibinfo{journal}{Phys. Rev. B} \textbf{\bibinfo{volume}{81}},
  \bibinfo{pages}{081411} (\bibinfo{year}{2010}).

\bibitem[{\citenamefont{Ladd et~al.}(1986)\citenamefont{Ladd, Moran, and
  Hoover}}]{Ladd_1986}
\bibinfo{author}{\bibfnamefont{A.~J.~C.} \bibnamefont{Ladd}},
  \bibinfo{author}{\bibfnamefont{B.}~\bibnamefont{Moran}}, \bibnamefont{and}
  \bibinfo{author}{\bibfnamefont{W.~G.} \bibnamefont{Hoover}},
  \bibinfo{journal}{Phys. Rev. B} \textbf{\bibinfo{volume}{34}},
  \bibinfo{pages}{5058} (\bibinfo{year}{1986}).

\bibitem[{\citenamefont{McGaughey and Kaviany}(2004)}]{McGaughey_2004}
\bibinfo{author}{\bibfnamefont{A.~J.~H.} \bibnamefont{McGaughey}}
  \bibnamefont{and} \bibinfo{author}{\bibfnamefont{M.}~\bibnamefont{Kaviany}},
  \bibinfo{journal}{Phys. Rev. B.} \textbf{\bibinfo{volume}{69}},
  \bibinfo{pages}{094303} (\bibinfo{year}{2004}).

\bibitem[{\citenamefont{Turney et~al.}(2009{\natexlab{a}})\citenamefont{Turney,
  Landry, McGaughey, and Amon}}]{Turney_2009}
\bibinfo{author}{\bibfnamefont{J.~E.} \bibnamefont{Turney}},
  \bibinfo{author}{\bibfnamefont{E.~S.} \bibnamefont{Landry}},
  \bibinfo{author}{\bibfnamefont{A.~J.~H.} \bibnamefont{McGaughey}},
  \bibnamefont{and} \bibinfo{author}{\bibfnamefont{C.~H.} \bibnamefont{Amon}},
  \bibinfo{journal}{Phys. Rev. B} \textbf{\bibinfo{volume}{79}},
  \bibinfo{pages}{064301} (\bibinfo{year}{2009}{\natexlab{a}}).

\bibitem[{\citenamefont{Henry and Chen}(2008)}]{Henry_2008}
\bibinfo{author}{\bibfnamefont{A.~S.} \bibnamefont{Henry}} \bibnamefont{and}
  \bibinfo{author}{\bibfnamefont{G.}~\bibnamefont{Chen}}, \bibinfo{journal}{J.
  Comput. Theor. Nanosci.} \textbf{\bibinfo{volume}{5}}, \bibinfo{pages}{1}
  (\bibinfo{year}{2008}).

\bibitem[{\citenamefont{de~Koker}(2009)}]{Koker_2009}
\bibinfo{author}{\bibfnamefont{N.}~\bibnamefont{de~Koker}},
  \bibinfo{journal}{Phys. Rev. Lett.} \textbf{\bibinfo{volume}{103}},
  \bibinfo{pages}{125902} (\bibinfo{year}{2009}).

\bibitem[{\citenamefont{Shiomi and Maruyama}(2006)}]{Shiomi_2006}
\bibinfo{author}{\bibfnamefont{J.}~\bibnamefont{Shiomi}} \bibnamefont{and}
  \bibinfo{author}{\bibfnamefont{S.}~\bibnamefont{Maruyama}},
  \bibinfo{journal}{Phys. Rev. B} \textbf{\bibinfo{volume}{73}},
  \bibinfo{pages}{205420} (\bibinfo{year}{2006}).

\bibitem[{\citenamefont{Qiu et~al.}(2011{\natexlab{a}})\citenamefont{Qiu, Bao,
  Zhang, Wu, and Ruan}}]{Qiu_2011_pbte}
\bibinfo{author}{\bibfnamefont{B.}~\bibnamefont{Qiu}},
  \bibinfo{author}{\bibfnamefont{H.}~\bibnamefont{Bao}},
  \bibinfo{author}{\bibfnamefont{G.}~\bibnamefont{Zhang}},
  \bibinfo{author}{\bibfnamefont{Y.}~\bibnamefont{Wu}}, \bibnamefont{and}
  \bibinfo{author}{\bibfnamefont{X.}~\bibnamefont{Ruan}},
  \bibinfo{journal}{Comput. Mater. Sci.} \textbf{\bibinfo{volume}{53}},
  \bibinfo{pages}{278} (\bibinfo{year}{2011}{\natexlab{a}}).

\bibitem[{\citenamefont{Qiu et~al.}(2011{\natexlab{b}})\citenamefont{Qiu, Wang,
  and Ruan}}]{Qiu_2011_mrs}
\bibinfo{author}{\bibfnamefont{B.}~\bibnamefont{Qiu}},
  \bibinfo{author}{\bibfnamefont{Y.}~\bibnamefont{Wang}}, \bibnamefont{and}
  \bibinfo{author}{\bibfnamefont{X.}~\bibnamefont{Ruan}},
  \bibinfo{journal}{Proceedings of MRS spring meeting} pp.
  \bibinfo{pages}{MRSS11--1347--BB10--08} (\bibinfo{year}{2011}{\natexlab{b}}).

\bibitem[{\citenamefont{Qiu and Ruan}(2011)}]{Qiu_2011_imece}
\bibinfo{author}{\bibfnamefont{B.}~\bibnamefont{Qiu}} \bibnamefont{and}
  \bibinfo{author}{\bibfnamefont{X.}~\bibnamefont{Ruan}},
  \bibinfo{journal}{Proceedings of ASME IMECE 2011} pp.
  \bibinfo{pages}{IMECE2011--62963} (\bibinfo{year}{2011}).

\bibitem[{\citenamefont{Ong et~al.}(2011)\citenamefont{Ong, Pop, and
  Shiomi}}]{Ong_2011_cnt}
\bibinfo{author}{\bibfnamefont{Z.-Y.} \bibnamefont{Ong}},
  \bibinfo{author}{\bibfnamefont{E.}~\bibnamefont{Pop}}, \bibnamefont{and}
  \bibinfo{author}{\bibfnamefont{J.}~\bibnamefont{Shiomi}},
  \bibinfo{journal}{Phys. Rev. B} \textbf{\bibinfo{volume}{84}},
  \bibinfo{pages}{165418} (\bibinfo{year}{2011}).

\bibitem[{\citenamefont{Larkin et~al.}(2012)\citenamefont{Larkin, Massicotte,
  Turney, and McGaughey}}]{Larkin_2012}
\bibinfo{author}{\bibfnamefont{J.}~\bibnamefont{Larkin}},
  \bibinfo{author}{\bibfnamefont{A.~D.} \bibnamefont{Massicotte}},
  \bibinfo{author}{\bibfnamefont{J.~E.} \bibnamefont{Turney}},
  \bibnamefont{and} \bibinfo{author}{\bibfnamefont{A.~J.~H.}
  \bibnamefont{McGaughey}}, \bibinfo{journal}{ASME Micro/Nanoscale Heat and
  Mass Transfer International Conference}  (\bibinfo{year}{2012}).

\bibitem[{\citenamefont{Dove}(1993)}]{Dove_1993}
\bibinfo{author}{\bibfnamefont{M.~T.} \bibnamefont{Dove}},
  \emph{\bibinfo{title}{Introduction to lattice dynamics}}
  (\bibinfo{publisher}{Cambridge University Press},
  \bibinfo{address}{Cambridge}, \bibinfo{year}{1993}).

\bibitem[{\citenamefont{Henry and Chen}(2009)}]{Henry_2009}
\bibinfo{author}{\bibfnamefont{A.~S.} \bibnamefont{Henry}} \bibnamefont{and}
  \bibinfo{author}{\bibfnamefont{G.}~\bibnamefont{Chen}},
  \bibinfo{journal}{Phys. Rev. B} \textbf{\bibinfo{volume}{79}},
  \bibinfo{pages}{144305} (\bibinfo{year}{2009}).

\bibitem[{\citenamefont{Singh et~al.}(2011)\citenamefont{Singh, Murthy, and
  Fisher}}]{Singh_2011}
\bibinfo{author}{\bibfnamefont{D.}~\bibnamefont{Singh}},
  \bibinfo{author}{\bibfnamefont{J.~Y.} \bibnamefont{Murthy}},
  \bibnamefont{and} \bibinfo{author}{\bibfnamefont{T.~S.}
  \bibnamefont{Fisher}}, \bibinfo{journal}{arXiv:1104.4964v1}
  (\bibinfo{year}{2011}).

\bibitem[{\citenamefont{McQuarrie}(2000)}]{McQuarrie_2000}
\bibinfo{author}{\bibfnamefont{D.~A.} \bibnamefont{McQuarrie}},
  \emph{\bibinfo{title}{Statistical Mechanics}} (\bibinfo{publisher}{University
  Science Books}, \bibinfo{address}{Sausalito}, \bibinfo{year}{2000}).

\bibitem[{\citenamefont{Ashcroft and Mermin}(1976)}]{Ashcroft_1976}
\bibinfo{author}{\bibfnamefont{N.~W.} \bibnamefont{Ashcroft}} \bibnamefont{and}
  \bibinfo{author}{\bibfnamefont{N.~D.} \bibnamefont{Mermin}},
  \emph{\bibinfo{title}{Solid State Physics}} (\bibinfo{publisher}{Brooks
  Cole}, \bibinfo{year}{1976}).

\bibitem[{\citenamefont{Wang et~al.}(2008)\citenamefont{Wang, Lim, Ng,
  Ozyilmaz, and Kuok}}]{Wang_2008}
\bibinfo{author}{\bibfnamefont{Z.~K.} \bibnamefont{Wang}},
  \bibinfo{author}{\bibfnamefont{H.~S.} \bibnamefont{Lim}},
  \bibinfo{author}{\bibfnamefont{S.~C.} \bibnamefont{Ng}},
  \bibinfo{author}{\bibfnamefont{B.}~\bibnamefont{Ozyilmaz}}, \bibnamefont{and}
  \bibinfo{author}{\bibfnamefont{M.~H.} \bibnamefont{Kuok}},
  \bibinfo{journal}{Carbon} \textbf{\bibinfo{volume}{46}},
  \bibinfo{pages}{2133} (\bibinfo{year}{2008}).

\bibitem[{\citenamefont{Wang et~al.}(2011)\citenamefont{Wang, Xie, Bui, Liu,
  Ni, Li, and Thong}}]{Wang_2011}
\bibinfo{author}{\bibfnamefont{Z.}~\bibnamefont{Wang}},
  \bibinfo{author}{\bibfnamefont{R.}~\bibnamefont{Xie}},
  \bibinfo{author}{\bibfnamefont{C.~T.} \bibnamefont{Bui}},
  \bibinfo{author}{\bibfnamefont{D.}~\bibnamefont{Liu}},
  \bibinfo{author}{\bibfnamefont{X.}~\bibnamefont{Ni}},
  \bibinfo{author}{\bibfnamefont{B.}~\bibnamefont{Li}}, \bibnamefont{and}
  \bibinfo{author}{\bibfnamefont{J.~T.~L.} \bibnamefont{Thong}},
  \bibinfo{journal}{Nano Lett.} \textbf{\bibinfo{volume}{11}},
  \bibinfo{pages}{113} (\bibinfo{year}{2011}).

\bibitem[{\citenamefont{Kang et~al.}(2010)\citenamefont{Kang, Abdula, Cahill,
  and Shim}}]{Kang_2010}
\bibinfo{author}{\bibfnamefont{K.}~\bibnamefont{Kang}},
  \bibinfo{author}{\bibfnamefont{D.}~\bibnamefont{Abdula}},
  \bibinfo{author}{\bibfnamefont{D.~G.} \bibnamefont{Cahill}},
  \bibnamefont{and} \bibinfo{author}{\bibfnamefont{M.}~\bibnamefont{Shim}},
  \bibinfo{journal}{Phys. Rev. B} \textbf{\bibinfo{volume}{81}},
  \bibinfo{pages}{165405} (\bibinfo{year}{2010}).

\bibitem[{\citenamefont{Wang et~al.}(2010)\citenamefont{Wang, Strait, George,
  Shivaraman, Shields, Chandrashekhar, Hwang, Rana, Spencer, Ruiz-Vargas
  et~al.}}]{Wang_2010_ram}
\bibinfo{author}{\bibfnamefont{H.}~\bibnamefont{Wang}},
  \bibinfo{author}{\bibfnamefont{J.~H.} \bibnamefont{Strait}},
  \bibinfo{author}{\bibfnamefont{P.~A.} \bibnamefont{George}},
  \bibinfo{author}{\bibfnamefont{S.}~\bibnamefont{Shivaraman}},
  \bibinfo{author}{\bibfnamefont{V.~B.} \bibnamefont{Shields}},
  \bibinfo{author}{\bibfnamefont{M.}~\bibnamefont{Chandrashekhar}},
  \bibinfo{author}{\bibfnamefont{J.}~\bibnamefont{Hwang}},
  \bibinfo{author}{\bibfnamefont{F.}~\bibnamefont{Rana}},
  \bibinfo{author}{\bibfnamefont{M.~G.} \bibnamefont{Spencer}},
  \bibinfo{author}{\bibfnamefont{C.~S.} \bibnamefont{Ruiz-Vargas}},
  \bibnamefont{et~al.}, \bibinfo{journal}{Appl. Phys. Lett.}
  \textbf{\bibinfo{volume}{96}}, \bibinfo{pages}{081917}
  (\bibinfo{year}{2010}).

\bibitem[{\citenamefont{Calizo et~al.}(2007)\citenamefont{Calizo, Bao, Miao,
  Lau, and Balandin}}]{Calizo_2007}
\bibinfo{author}{\bibfnamefont{I.}~\bibnamefont{Calizo}},
  \bibinfo{author}{\bibfnamefont{W.}~\bibnamefont{Bao}},
  \bibinfo{author}{\bibfnamefont{F.}~\bibnamefont{Miao}},
  \bibinfo{author}{\bibfnamefont{C.~N.} \bibnamefont{Lau}}, \bibnamefont{and}
  \bibinfo{author}{\bibfnamefont{A.~A.} \bibnamefont{Balandin}},
  \bibinfo{journal}{Appl. Phys. Lett.} \textbf{\bibinfo{volume}{91}},
  \bibinfo{pages}{201904} (\bibinfo{year}{2007}).

\bibitem[{\citenamefont{Faugeras et~al.}(2010)\citenamefont{Faugeras, Faugeras,
  Orlita, Potemski, Nair, and Geim}}]{Faugeras_2010}
\bibinfo{author}{\bibfnamefont{C.}~\bibnamefont{Faugeras}},
  \bibinfo{author}{\bibfnamefont{B.}~\bibnamefont{Faugeras}},
  \bibinfo{author}{\bibfnamefont{M.}~\bibnamefont{Orlita}},
  \bibinfo{author}{\bibfnamefont{M.}~\bibnamefont{Potemski}},
  \bibinfo{author}{\bibfnamefont{R.~R.} \bibnamefont{Nair}}, \bibnamefont{and}
  \bibinfo{author}{\bibfnamefont{A.~K.} \bibnamefont{Geim}},
  \bibinfo{journal}{ACS Nano} \textbf{\bibinfo{volume}{4}},
  \bibinfo{pages}{1889} (\bibinfo{year}{2010}).

\bibitem[{\citenamefont{Mattausch and Pankratov}(2007)}]{Mattausch_2007}
\bibinfo{author}{\bibfnamefont{A.}~\bibnamefont{Mattausch}} \bibnamefont{and}
  \bibinfo{author}{\bibfnamefont{O.}~\bibnamefont{Pankratov}},
  \bibinfo{journal}{Phys. Rev. Lett.} \textbf{\bibinfo{volume}{99}},
  \bibinfo{pages}{076802} (\bibinfo{year}{2007}).

\bibitem[{\citenamefont{Pettes et~al.}(2011)\citenamefont{Pettes, Jo, Yao, and
  Shi}}]{Pettes_2011}
\bibinfo{author}{\bibfnamefont{M.~T.} \bibnamefont{Pettes}},
  \bibinfo{author}{\bibfnamefont{I.}~\bibnamefont{Jo}},
  \bibinfo{author}{\bibfnamefont{Z.}~\bibnamefont{Yao}}, \bibnamefont{and}
  \bibinfo{author}{\bibfnamefont{L.}~\bibnamefont{Shi}}, \bibinfo{journal}{Nano
  Lett.} \textbf{\bibinfo{volume}{11}}, \bibinfo{pages}{1195}
  (\bibinfo{year}{2011}).

\bibitem[{\citenamefont{Brenner et~al.}(2002)\citenamefont{Brenner, Shenderova,
  Harrison, Stuart, Ni, and Sinnott}}]{Brenner_2002}
\bibinfo{author}{\bibfnamefont{D.~W.} \bibnamefont{Brenner}},
  \bibinfo{author}{\bibfnamefont{O.~A.} \bibnamefont{Shenderova}},
  \bibinfo{author}{\bibfnamefont{J.~A.} \bibnamefont{Harrison}},
  \bibinfo{author}{\bibfnamefont{S.~J.} \bibnamefont{Stuart}},
  \bibinfo{author}{\bibfnamefont{B.}~\bibnamefont{Ni}}, \bibnamefont{and}
  \bibinfo{author}{\bibfnamefont{S.~B.} \bibnamefont{Sinnott}},
  \bibinfo{journal}{J Physics: Condensed Matter} \textbf{\bibinfo{volume}{14}},
  \bibinfo{pages}{783} (\bibinfo{year}{2002}).

\bibitem[{\citenamefont{Turney et~al.}(2009{\natexlab{b}})\citenamefont{Turney,
  McGaughey, and Amon}}]{Turney_2009_qc}
\bibinfo{author}{\bibfnamefont{J.~E.} \bibnamefont{Turney}},
  \bibinfo{author}{\bibfnamefont{A.~J.~H.} \bibnamefont{McGaughey}},
  \bibnamefont{and} \bibinfo{author}{\bibfnamefont{C.~H.} \bibnamefont{Amon}},
  \bibinfo{journal}{Phys. Rev. B} \textbf{\bibinfo{volume}{79}},
  \bibinfo{pages}{224305} (\bibinfo{year}{2009}{\natexlab{b}}).

\end{thebibliography}

\end{document}